\documentclass[sigconf,9pt]{acmart}
\AtBeginDocument{%
  }

\copyrightyear{2024}
\acmYear{2024}
\setcopyright{acmlicensed}\acmConference[ICCAD '24]{IEEE/ACM International Conference on Computer-Aided Design}{October 27--31, 2024}{New York, NY, USA}
\acmBooktitle{IEEE/ACM International Conference on Computer-Aided Design (ICCAD '24), October 27--31, 2024, New York, NY, USA}
\acmDOI{10.1145/3676536.3676687}
\acmISBN{979-8-4007-1077-3/24/10}

\usepackage{xcolor}

\usepackage{graphicx}

\usepackage{tikz}
\usepackage[most]{tcolorbox}
\usepackage{gensymb}
\usepackage{todonotes}
\usepackage{amsmath,amsfonts}
\usepackage{algorithm}
\usepackage[noend]{algpseudocode}
\usepackage[symbol]{footmisc}

\usepackage{tikz}
\usepackage{amsmath}
\usepackage{multirow}
\usepackage{booktabs}

\newcommand*\circled[1]{\tikz[baseline=(char.base)]{
            \node[shape=circle,fill,inner sep=1pt] (char) {\textcolor{white}{#1}};}}

\usepackage{xcolor}
\usepackage{pgfplots}
\usepackage{tikz}
\pgfplotsset{compat=1.8}
\pgfplotsset{
    width=\textwidth,
}
\usepackage{lipsum}
\usepackage{pgfplotstable}
\usetikzlibrary{pgfplots.groupplots}

\usepackage{tikz}

\definecolor{codegreen}{rgb}{0,0.6,0}
\definecolor{codegray}{rgb}{0.5,0.5,0.5}
\definecolor{codepurple}{rgb}{0.58,0,0.82}
\definecolor{mGreen}{rgb}{0,0.6,0}
\definecolor{mGray}{rgb}{0.5,0.5,0.5}
\definecolor{mPurple}{rgb}{0.58,0,0.82}
\definecolor{backcolour}{rgb}{0.95,0.95,0.92}

\definecolor{RYB1}{RGB}{80, 99, 42}
\definecolor{RYB2}{RGB}{215, 227, 191}
\definecolor{RYB3}{RGB}{198, 187, 174}
\definecolor{RYB4}{RGB}{146, 205, 220}
\definecolor{RYB5}{RGB}{238, 144, 34}
\definecolor{RYB6}{RGB}{142, 172, 59}

\pgfplotscreateplotcyclelist{colorbrewer-RYB}{
{RYB1!50!black,fill=RYB1},
{RYB2!50!black,fill=RYB2},
{RYB3!50!black,fill=RYB3},
{RYB4!50!black,fill=RYB4},
{RYB5!50!black,fill=RYB5},
{RYB6!50!black,fill=RYB6},
}

\usepackage{lscape} 
\usepackage{pdflscape}
\usepackage{afterpage}
\usepackage{capt-of}
\usepackage{listings}
\usepackage{float}
\usepackage{array,multirow,graphicx}
\usepackage{caption}
\usepackage{subcaption}
\usepackage{soul}
\usepackage{pifont}

\definecolor{ggreen}{HTML}{2CC225}
\definecolor{yyellow}{HTML}{C2C80A}
\definecolor{bbrown}{HTML}{8e4603}

\definecolor{codegreen}{rgb}{0,0.6,0}
\definecolor{codegray}{rgb}{0.5,0.5,0.5}
\definecolor{codepurple}{rgb}{0.58,0,0.82}
\definecolor{mGreen}{rgb}{0,0.6,0}
\definecolor{mGray}{rgb}{0.5,0.5,0.5}
\definecolor{mPurple}{rgb}{0.58,0,0.82}
\definecolor{backcolour}{rgb}{0.95,0.95,0.92}

\lstdefinestyle{CStyle}{
    commentstyle=\color{mGreen},
    keywordstyle=\color{magenta},
    numberstyle=\tiny\color{mGray},
    stringstyle=\color{mPurple},
    basicstyle=\sffamily\footnotesize,
    frame=lrtb,
    breakatwhitespace=false,         
    breaklines=true,                 
    captionpos=b,                    
    keepspaces=true,                 
    numbers=left,                    
    numbersep=5pt,                  
    showspaces=false,                
    showstringspaces=false,
    showtabs=false,                  
    tabsize=2,
    language=C
}

\lstdefinestyle{CStyle1}{
    commentstyle=\color{mGreen},
    keywordstyle=\color{magenta},
    numberstyle=\tiny\color{mGray},
    stringstyle=\color{mPurple},
    basicstyle=\sffamily\footnotesize,    frame=lrtb,
    breakatwhitespace=false,         
    breaklines=true,                 
    captionpos=b,                    
    keepspaces=true,                 
    numbers=left,                    
    numbersep=5pt,                  
    showspaces=false,                
    showstringspaces=false,
    showtabs=false,                  
    tabsize=2,
    language=C
}

\lstdefinestyle{mystyle}{
    commentstyle=\color{codegreen},
    keywordstyle=\color{magenta},
    numberstyle=\tiny\color{codegray},
    stringstyle=\color{codepurple},
    basicstyle=\sffamily\footnotesize,
    breakatwhitespace=false,         
    breaklines=true,                 
    captionpos=b,                    
    keepspaces=true,                 
    numbers=left,                    
    numbersep=5pt,                  
    showspaces=false,                
    showstringspaces=false,
    showtabs=false,                  
    tabsize=2,
    language=C
}
\lstdefinestyle{trans}{
    commentstyle=\color{codegray},
    numberstyle=\tiny\color{codegray},
    stringstyle=\color{codepurple},
     basicstyle=\sffamily\footnotesize,
    frame=lrtb,
    breakatwhitespace=false,         
    breaklines=true,                 
    captionpos=b,                    
    keepspaces=true,                 
    numbers=left,                    
    numbersep=5pt,                  
    showspaces=false,                
    showstringspaces=false,
    showtabs=false,                  
    tabsize=2,
     language=[x86masm]Assembler,  escapeinside={\%*}{*)},   
     }     

\begin{document}


\title{RandOhm: Mitigating Impedance Side-channel Attacks using Randomized Circuit Configurations}

\author{Saleh Khalaj Monfared}
\affiliation{%
  \institution{Worcester Polytechnic Institute}
     \city{Worcester}
  \state{MA}
    \country{USA}
}
\email{skmonfared@wpi.edu}

\author{Domenic Forte}
\affiliation{%
  \institution{University of Florida}
       \city{Gainesville}
        \state{FL}
        \country{USA}
 }
\email{dforte@ece.ufl.edu}

\author{Shahin Tajik}
\affiliation{%
  \institution{Worcester Polytechnic Institute}
     \city{Worcester}
  \state{MA}
    \country{USA}
  }
\email{stajik@wpi.edu}

\renewcommand{\shortauthors}{Khalaj Monfared et al.}

\begin{abstract}
  
Physical side-channel attacks can compromise the security of integrated circuits. Most physical side-channel attacks (e.g., power or electromagnetic) exploit the dynamic behavior of a chip, typically manifesting as changes in current consumption or voltage fluctuations where algorithmic countermeasures, such as masking, can effectively mitigate them. However, as demonstrated recently, these mitigation techniques are not entirely effective against backscattered side-channel attacks such as impedance analysis. In the case of an impedance attack, an adversary exploits the data-dependent impedance variations of the chip’s power delivery network (PDN) to extract secret information. In this work, we introduce \emph{RandOhm}, which exploits a moving target defense (MTD) strategy based on the partial reconfiguration (PR) feature of mainstream FPGAs and programmable SoCs to defend against impedance side-channel attacks. 
We demonstrate that the information leakage through the PDN’s impedance could be significantly reduced via runtime reconfiguration of the secret-sensitive parts of the circuitry. Hence, by constantly randomizing the placement and routing of the circuit, one can decorrelate the data-dependent computation from the impedance value.
Moreover, in contrast to existing PR-based countermeasures, \emph{RandOhm} deploys open-source bitstream manipulation tools on programmable SoCs to speed up the randomization and provide real-time protection.
To validate our claims, we apply \emph{RandOhm} to AES ciphers realized on 28-nm FPGAs. 
We analyze the resiliency of our approach by performing non-profiled and profiled impedance analysis attacks and investigate the overhead of our mitigation in terms of delay and performance.
\end{abstract}




\maketitle

 \section{Introduction}\label{sec:introduction}

\begin{figure*}[t!]
           \centering
        \begin{subfigure}{.52\linewidth}
         \includegraphics [width=\textwidth]{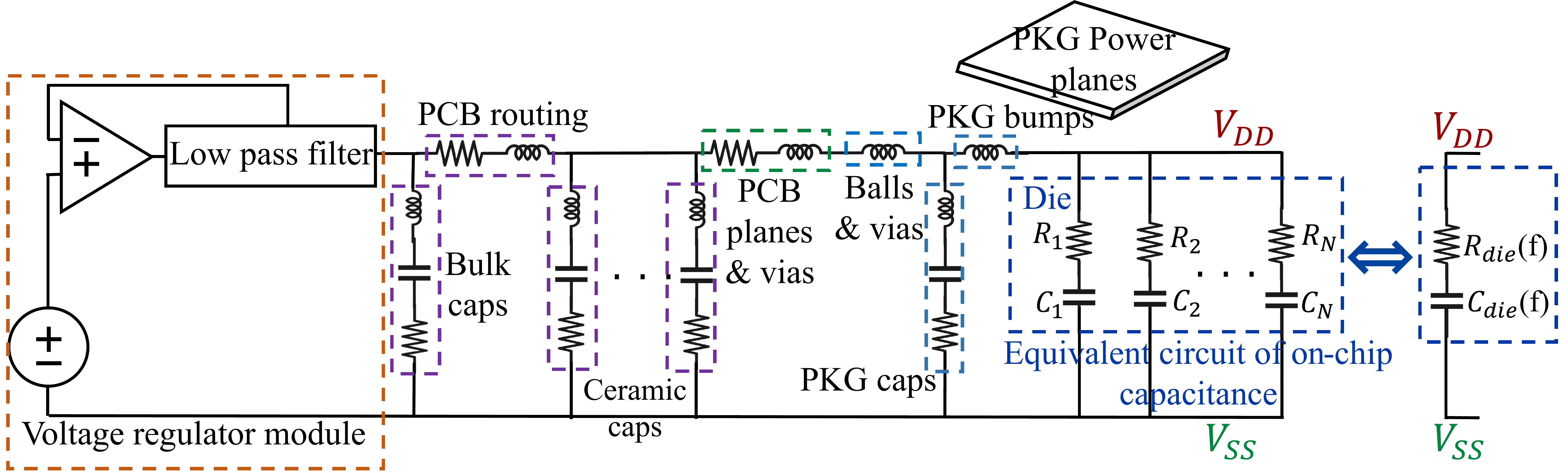}
                \caption{}
               \label{subfig:RLC_chip}
        \end{subfigure}
        \begin{subfigure}{.21\linewidth}
         \includegraphics [width=\textwidth]{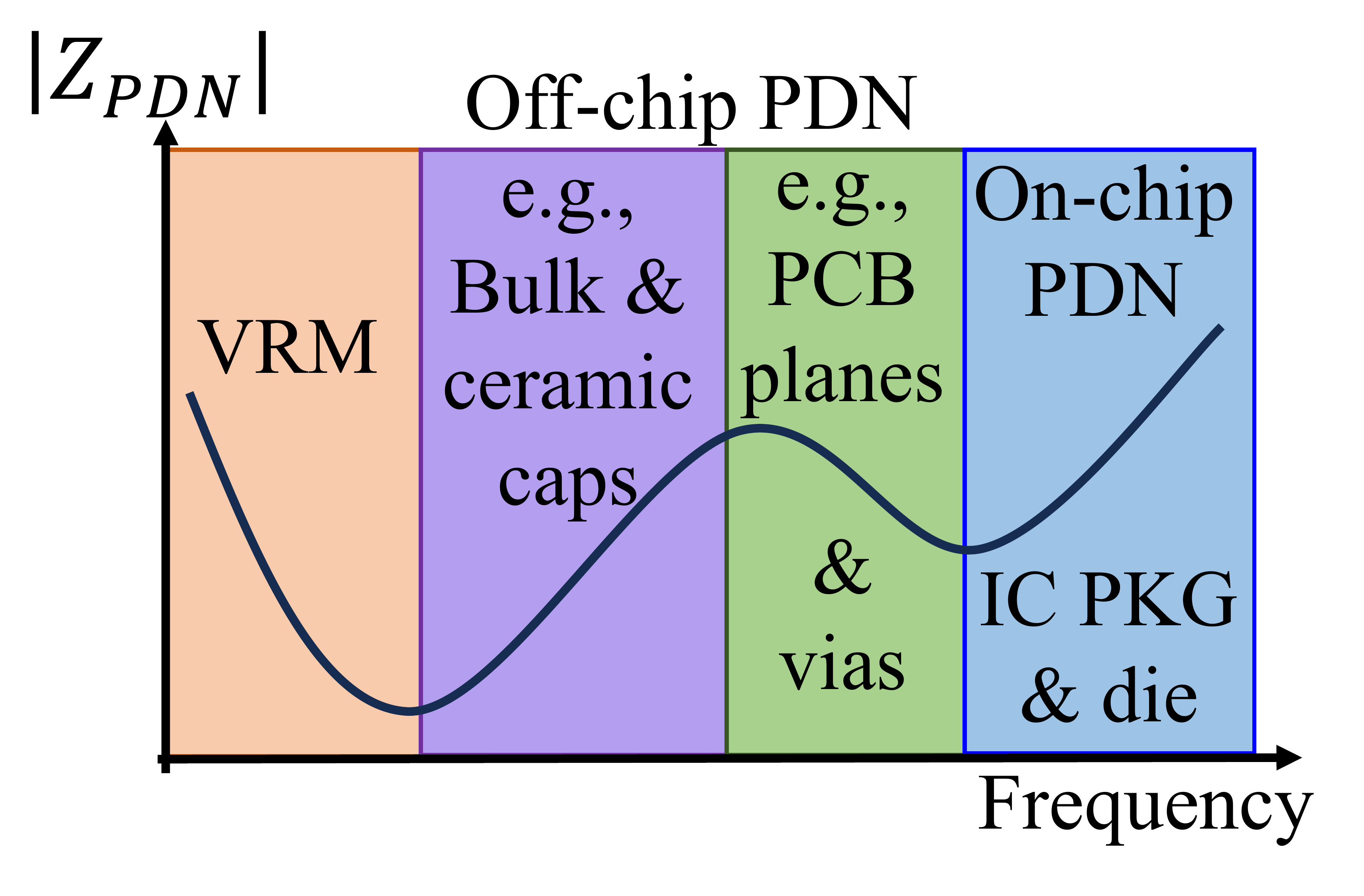}
                \caption{}
               \label{subfig:freq_disp}
                 \end{subfigure}
        \begin{subfigure}{.26\linewidth}
         \includegraphics [width=\textwidth]{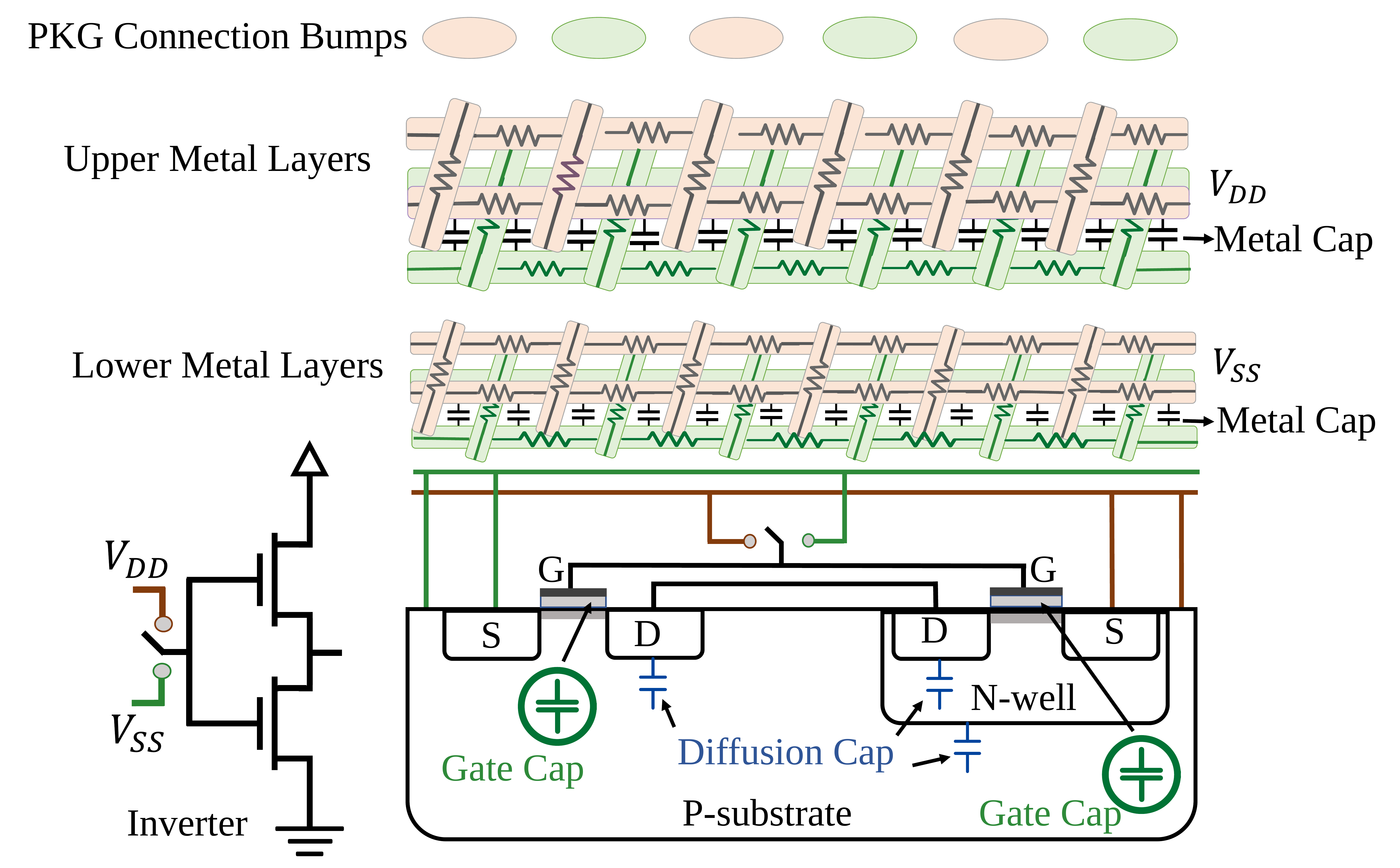}
                \caption{}
              \label{subfig:SourceOfchange}
                 \end{subfigure}
                 \vspace*{-5mm}

	\caption{(a) Equivalent RLC circuit model of the power distribution network (PDN) of the PCB and chip~\cite{monfared2023leakyohm}. (b) Contribution of different parts of the PDN to the impedance over frequency. (c) Contribution of a CMOS inverter to PDN's impedance~\cite{mosavirik2023silicon}.}
   	\label{fig:RLC_Model_freq_disp}
 \end{figure*}
 
Side-channel vulnerabilities can compromise the security of cryptographic implementations on integrated circuits (ICs). These vulnerabilities arise from the inherent effects of computation and data storage on factors like current consumption and supply voltage fluctuations within the IC. Such fluctuations manifest in various measurable ways, including power consumption~\cite{kocher1999differential}, electromagnetic (EM) emanation~\cite{vuagnoux2009compromising}, acoustic waves~\cite{backes2010acoustic}, photon emission~\cite{tajik2017power}, and thermal radiation~\cite{hutter2014temperature}. These characteristics have been exploited in various types of side-channel analysis (SCA) attacks to breach the security of diverse cryptographic implementations. 
Over the last two decades, various countermeasures have been developed to defeat these attacks.

However, the security of the chip has shown to be still vulnerable to a novel class of physical attacks known as \emph{active sensing} or \emph{backscattered} SCAs.
In such SCAs, the attacker stimulates the target device using signals in various forms, e.g., microwave radiations~\cite{monfared2023leakyohm,kaji2023echo,awal2023disassembling}, near-infrared laser beams~\cite{tajik2017power,krachenfels2021real,krachenfels2021automatic}, or even electron beams~\cite{amini2023electrons}, and measures the reflected/scattered signals from it.
The reflected/scattered signals are modulated depending on the state of a circuit or memory contents, and thus, can be exploited by the attacker to extract secret information from the chip.
Among these active SCAs, non-invasive stimulation using microwave signals, through the system's power delivery network (PDN)~\cite{monfared2023leakyohm,awal2023disassembling} or over the air~\cite{kaji2023echo}, is the most threatening one due to its effectiveness and inexpensive nature.
The main reason behind the modulation of the reflected microwave signal is the data-dependent changes in the \emph{impedance} of the chip.
In contrast to most of the conventional SCA attacks, such as power and EM analysis, capturing data leakages only during state transitions, impedance analysis attacks \emph{enable the extraction of static data.}

Deploying data randomization in countermeasures, such as masking~\cite{nikova2006threshold,gross2016domain}, is a conventional method to mitigate passive SCA attacks, as it prevents the repetition and integration of the measurements over multiple clock cycles. 
However, randomness becomes ineffective if the adversary halts the circuit or probes the circuit between two clock cycles and recovers the entire state of the circuit using attacks such as impedance analysis~\cite{monfared2023leakyohm}.
Similar to masking, which randomizes the power consumption of the chip, one solution to mitigate the impedance leakage would be the randomization of the circuit's impedance by constantly changing the physical structure of the circuit.
Such a moving target defense (MTD) can be realized using the partial reconfiguration features of mainstream FPGAs and programmable SoCs, as changing the placement and routing of the circuit changes the circuit's impedance.
Driven by this fact, the following research questions are raised: \emph{(1) Does partial reconfiguration provide enough impedance randomness to resist impedance side-channel attacks? (2) Could such techniques be deployed in a scalable, modular, and efficient manner on top of a masked implementation of a given target cryptographic core?}

\noindent\textbf{Our Contribution.} To answer the above questions, we introduce \textit{RandOhm}, a new approach that utilizes an end-to-end modular MTD strategy based on the partial reconfiguration of FPGAs and programmable SoCs to mitigate impedance side-channel attacks.
Compared to existing reconfiguration-based mitigation methods, \textit{RandOhm} generates randomized partial bitstreams once during the design phase and deploys them during runtime using an open-source bitstream manipulation tool to expedite the process, improve memory utilization on the FPGA, and provide real-time protection.
By randomizing the placement and routing of circuitry through runtime reconfiguration of secret-sensitive parts, we can decorrelate data-dependent computation from impedance values, significantly reducing the information leakage through the PDN's impedance.
To show the effectiveness of our approach, we use \textit{RandOhm} on 28-nm FPGAs and SoCs to protect the AES cipher implementations.
We assess the resiliency of our proposed solution by performing non-profiled and profiled impedance analysis.
Finally, we investigate the overhead of our mitigation in terms of delay and performance.

\noindent\textbf{Source Code Availability.} We  publish the source code of \textit{RandOhm} in : \url{https://github.com/vernamlab/RandOhm}

\section{Technical Background}\label{sec:Background}


\subsection{Impedance Side-channel Attacks}\label{sec:scattersca}
The power delivery network (PDN) ensures a steady and low-noise voltage supply to the electronic components on the printed circuit board (PCB). 
It receives power from the voltage regulator module (VRM) and routes it through power rails to the chip. 
The PDN system can be represented by an equivalent circuit model, as shown in Fig.~\ref{subfig:RLC_chip}. 
It consists of both off-chip and on-chip components, including bulk capacitors, PCB routing traces, vias, package, and on-chip power planes. The impedance contribution of these individual components to the total PDN impedance varies across different frequency bands, see Fig.~\ref{subfig:freq_disp}.
At high frequencies, the on-chip capacitance ($C_{die}$) and resistance ($R_{die}$) dominate the on-chip impedance characteristics of the PDN.
As shown in Fig~\ref{subfig:SourceOfchange}, the state of each individual logic gate (here, an inverter) affects the PDN's impedance at certain frequency bands. 
Hence, measuring the impedance enables an adversary to observe data-dependent fluctuations of impedance and, thus, perform a side-channel attack~\cite{monfared2023leakyohm}.

The impedance of an electrical element is a frequency-dependent complex number $Z(f)$ and is represented with real and imaginary parts or in polar form $|Z|\angle \theta$. 
Impedance is measured at a set of different frequencies in a wide range. 
A common practice is to make use of vector network analyzers (VNAs) to perform scattering parameter (S-Matrix) measurements and extract impedance values.
To conduct a scattering parameter measurement, VNA is connected to the target device's PDN and a series of frequency points are set for the measurements. 
During the measurement, RF sine waves with specific power are generated in those frequency points and are injected into the PDN. 
At the same time, the reflected  signals are received by the VNA, and the relative amplitude and phase at each frequency point are recorded. 
$S_{11}$ describes the scattering reflection rate of the element.
In other words, it quantifies the portion of reflected RF waves ($S_{11}=V^-/V^+$). 
Upon measurement of $S_{11}$, a simple transformation can be used to extract the impedance profile using $Z_{DUT}=Z_0(1+S_{11})/(1-S_{11})$.

\vspace*{-5mm}

\subsection{Partial Reconfiguration as Side-Channel Countermeasure} 
Partial Reconfiguration (PR) is a feature that allows for dynamic modification of a portion of the FPGA while the rest of the system continues to operate uninterrupted. This capability not only enhances flexibility but also reduces power consumption and increases system adaptability to changing conditions or requirements. PR enables the FPGA to adapt itself without needing a complete system reboot, thereby ensuring continuous operation and efficiency~\cite{koch2012partial1}. Furthermore, the use of PR in FPGAs has been shown to be particularly useful in mission-critical applications where system downtime is not acceptable and in situations requiring real-time processing capabilities~\cite{koch2012partial}. The technology allows for the efficient use of FPGA resources, as it enables the reuse of the hardware for different functions at different times, which is especially beneficial in resource-constrained environments~\cite{vipin2018fpga}.

Several side-channel countermeasures~\cite{mentens2008power,guneysu2011generic,heyszl2012localized,moradi2013comprehensive,hettwer2019securing,bow2020side,khan2021moving} deploy PR to defeat power and electromagnetic (EM) analysis attacks.
These efforts merely utilize PR to introduce jitter (realized by delay) to defeat power side channels. 
Other approaches include relocation of the functions to defeat EM attacks.
The main drawback of some of these solutions (e.g.,~\cite{mentens2008power}) is the limited available number of randomized PRs, leading to a linear increase in the complexity of the attack.
Moreover, the partial bitstreams in these schemes have to be stored on external non-volatile memory and invoked during runtime, resulting in a very high overhead.

  \begin{figure*}[!t] 
\centering
\includegraphics[width=0.8\linewidth]{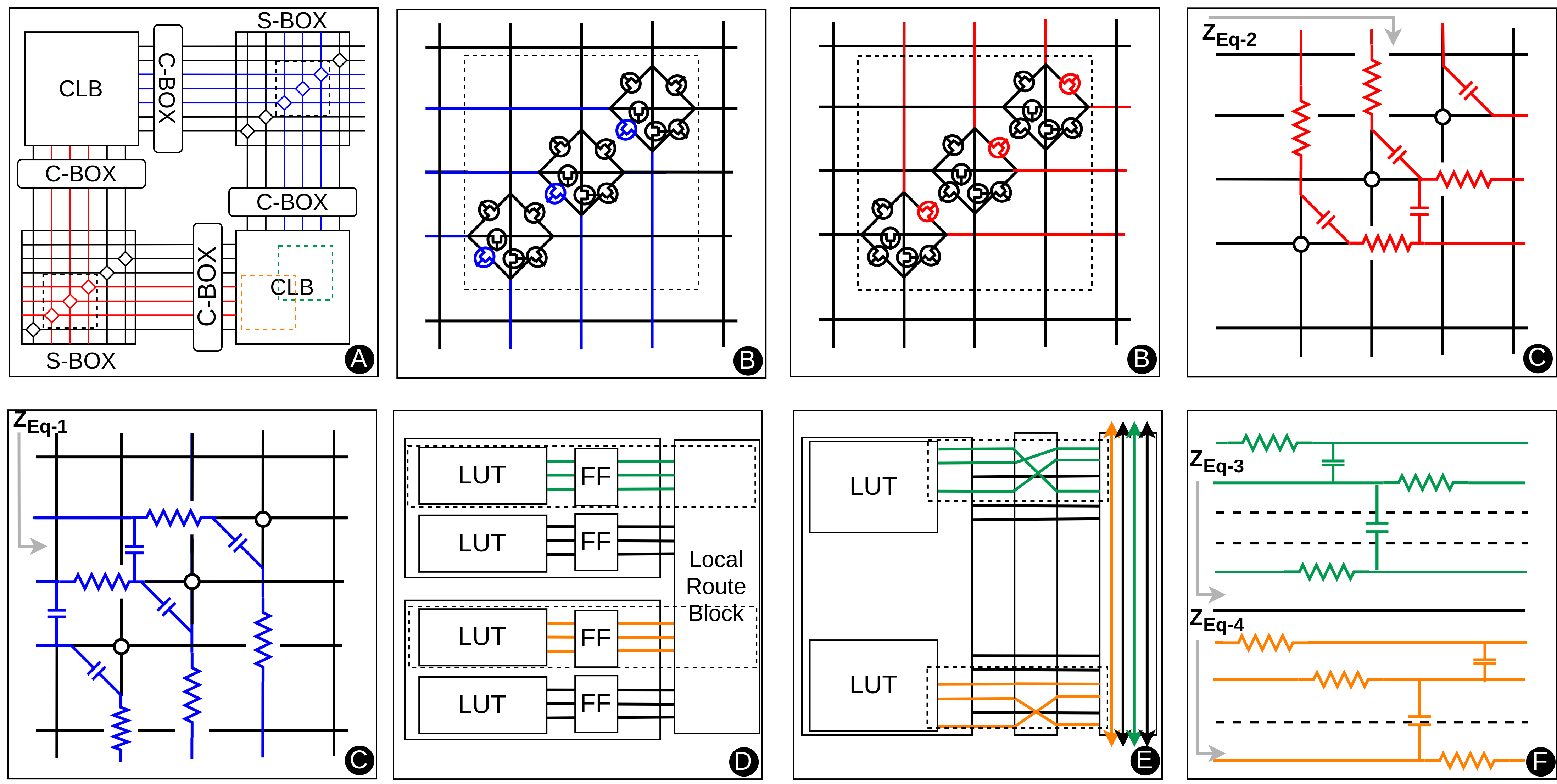}
\caption{Impact of different FPGA routing configurations on PDN's impedance}
\label{fig:clb}
\end{figure*}

\subsection{Bitstream Manipulation for Reconfiguration}\label{sec:pr}

AMD/Xilinx's Dynamic Function eXchange (DFX) introduces a method for defining PR regions within a static system, allowing users to assign modules to these regions on FPGA fabrics \cite{xilixreconfig}. 
However, there are several drawbacks in the Vivado toolchain, such as being too slow for real-time applications and the lack of support for bitstream relocation, limiting the maximum potential of reconfigurability~\cite{manev2022byteman}.
On the other hand, FPGA bitstream parsing and manipulation, which have been closely in touch with PR techniques, are thoroughly investigated by researchers~\cite{raghavan2002jpg}. BitMan~\cite{pham2017bitman} toolkit made it possible to relocate bitstream for several Xilinx devices. Other recent efforts, such as Bitfiltrator~\cite{kashani2022bitfiltrator}, strive to reverse engineer the bitstream encoding of the AMD/Xilinx FPGA families.
Recently, the open-source tool, known as Byteman~\cite{manev2022byteman}, improved the efficiency, speed, and compatibility of the existing bitstream manipulation tools by adding support for merging clock, CLB, BlockRAM data, and different merge strategies. More importantly, using such bitstream manipulation tools provides the ability to generate and deploy partial bitstreams of adjustable position and size without the help of proprietary slow toolchains, making it suitable for real-time applications.

\section{Hardware Multiplexing as MTD}\label{sec:multiplex}



\subsection{Dynamic Configurations and Impedance}
As shown in~\cite{mosavirik2023silicon}, the physical coordinates and its corresponding circuitry on the FPGA fabric leave distinguishable fingerprints on the PDN's impedance in the frequency domain, which can be exploited for mounting template attacks~\cite{monfared2023leakyohm}.
Fig.~\ref{fig:clb} depicts a series of high-level diagrams, each representing a specific part of the FPGA internals~\cite{amano2018principles}. Part \circled{A} in Fig.~\ref{fig:clb} assumes implementations of a particular function ($F$) in the bottom right Configurable Logic Block (CLB). 
Using different configurations, function $F$ can utilize orange \textcolor{orange}{SLICE 4} or green \textcolor{ggreen}{SLICE 3}. Furthermore, it is assumed that the routing to other CLBs could be implemented using either blue \textcolor{blue}{Switch-Box 1} or red \textcolor{red}{Switch-Box 2}. Regarding different routing configurations, parts \circled{B} illustrate the different activation of \textit{Switch-Boxes} routings. Based on each particular connection and state, the routing CMOS transistors and their simplified equivalent circuitry for \textcolor{blue}{Switch-Box 1} or red \textcolor{red}{Switch-Box 2} are shown in parts \circled{C}. As highlighted, the equivalent impedance seen from the PDN of the FPGA in each of these cases are different due to the differences in resistance and mutual capacitance for the wiring in each configuration ($Z_{Eq-1}$ and $Z_{Eq-2}$). Moreover, depending on the chosen slice indicated in part \circled{D}, a particular LUT (either orange \textcolor{orange}{SLICE 4} or green \textcolor{ggreen}{SLICE 3}) is selected. This leads to specific internal local connections at the transistor-level layout which is depicted in part \circled{E}. As illustrated in part \circled{F}, these configurations differ in terms of the equivalent impedance of $Z_{Eq-3}$ and $Z_{Eq-4}$ when different wirings are activated. It is worthy to highlight the geometrical asymmetry of physical elements on the die. This element-level asymmetry in the FPGA fabric yields a unique impedance for each implementation. 

Another observation is the possible importance of the measurement port. The estimated equivalent impedance using $S_{11}$ is often seen and measured from the PDN (and from specific ports on the chip). However, if the signals are injected and received from other physical ports (if applicable), new modes of physical asymmetry are achieved in terms of scattering parameters. 

We take these observations and explanations into account and design simple experiments to investigate the alterations in impedance profile by the use of multiplexing and run-time circuit modifications.


\subsection{Target Slice Multiplexer}
\begin{figure}[t!]
  \centering \noindent
   \includegraphics[width=0.85\linewidth]{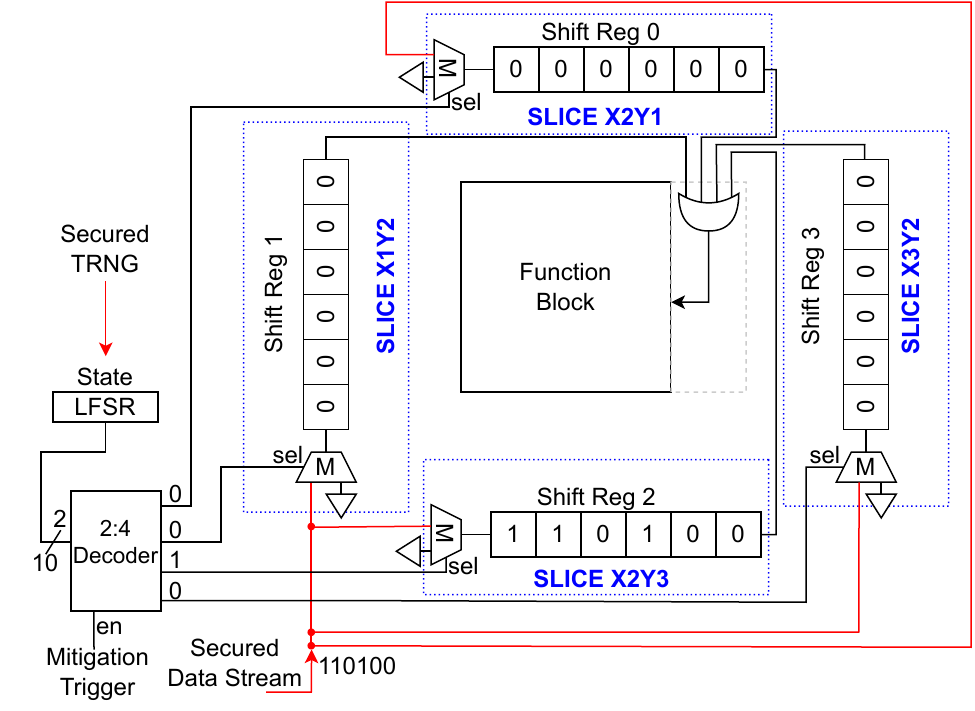}
	\caption{Real-time Target Slice Multiplexer}
            \label{fig:target_mul}                 
 \end{figure}

Although by exploiting the DFX, it is possible to entirely replace a module from one physical slice to another, a simple alternative is to have multiple instances of the same target circuit in distinct slices and choose one randomly to be connected periodically. The obvious trade-off here is the area overhead caused by all those additional blocks. However, as a simple mitigation, a real-time target slice multiplexer (see Fig.~\ref{fig:target_mul}) could be considered.

In Fig.~\ref{fig:target_mul}, with the assumption that the initial target data is securely stored and streamed into the functional block (highlighted with red), a linear feedback shift register (LFSR) is securely initialized with a TRNG and serves as a random selector for existing target shift-register in different physical slices on the FPGA (denoted by blue). Once the mechanism is activated, a single shift-register is chosen to load the data where other instances are cleared simultaneously. Furthermore, it is possible to re-activate the mechanism by including a trigger signal in the design.


 \subsection{Register Sequence Multiplexer}\label{seq}
\begin{figure}[t]
  \centering \noindent
   \includegraphics[width=0.65\linewidth]{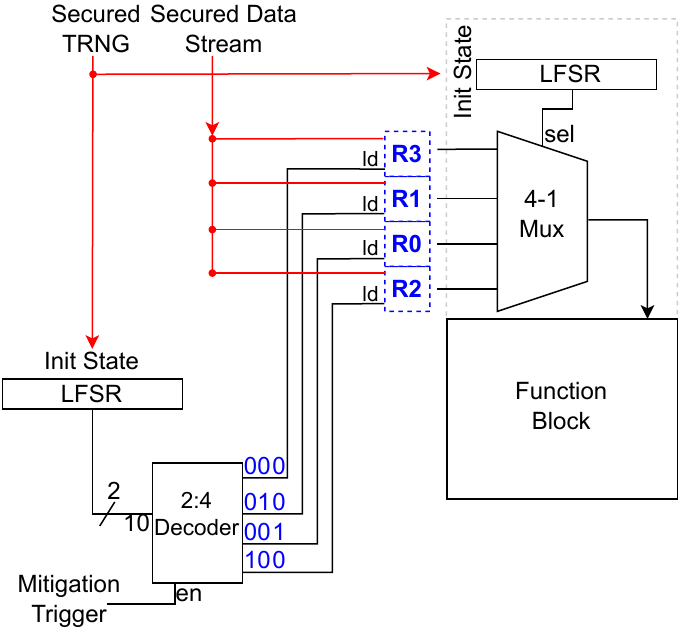}
	\caption{Real-time Register Sequence Multiplexer}
            \label{fig:reg_seq}                 
 \end{figure}

The slice multiplexing method presented earlier is a coarse-grained MTD and is confined within the available reconfigurable slices for the target data. Hence, the number of configurations could be up to hundreds, which is needed to tackle impedance attack scenarios. 
To surpass such limitations, we introduce a fine-grained MTD which involves hardware scrambling of register references. Fig.~\ref{fig:reg_seq} illustrates a simple digital design diagram of a real-time register sequence multiplexer. 

As depicted in Fig.~\ref{fig:reg_seq}, a securely initialized LFSR is used to determine a randomized sequence of a data load operation. This yields to randomization of the data order every time the target registers are loaded. The vital part of this mitigation is to maintain the initial state in order to read the data in the correct format by the function block. 
This procedure here could be considered as an inspiration from logic locking techniques~\cite{yasin2017evolution}. 
However, instead of locking the functionality of circuitry, we lock the sequence of a data load which leads to a considerable degree of MTD complexity. Theoretically, this method realizes the upper bound of super-exponential ($\mathcal{O}(n!)$) complexity against trial-based attacks.

 \section{Modular Moving Target Defense}\label{modular}

\subsection{High-Level Overview}\label{sec:highlevel}

\begin{figure*}[!t] 
\centering
\includegraphics[width=0.9\linewidth]{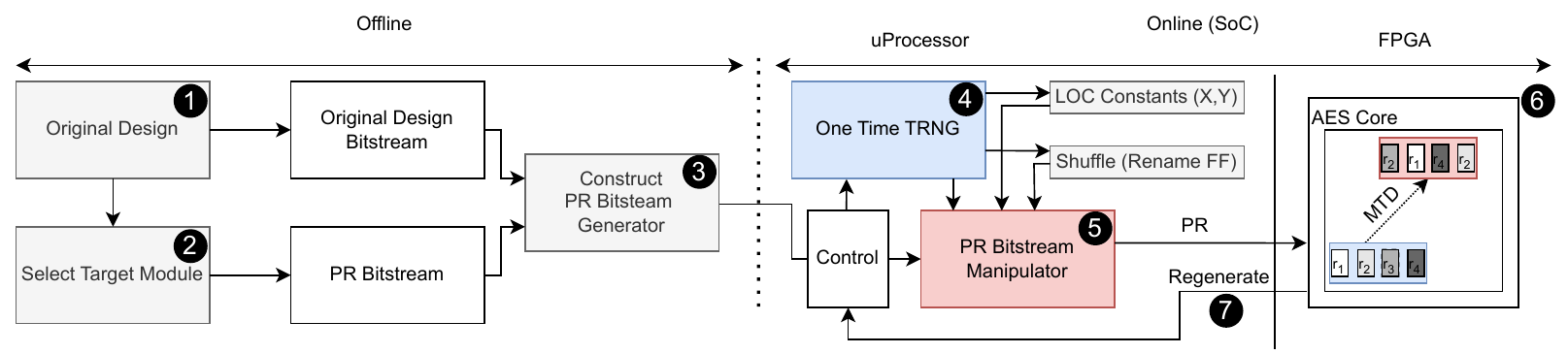}
\caption{High-level block-diagram description of \textit{RandOhm}}
\label{fig:overview}
\end{figure*}

Fig.~\ref{fig:overview} shows a block diagram of a high-level description of \textit{RandOhm}. 
The framework is divided into offline and online procedures. The offline part is executed once for a given target.
As indicated in \circled{1}, an original hardware design is considered. This design could be any core containing sensitive information that should be protected. For the sake of simplicity, we assume an AES algorithm as the original design for our descriptions. Using high-level scripting language (specifically TCL), in \circled{2}, targeted modules are indicated by the user. This is done by pre-defined annotations using a high-level scripting. Multiple partial bitstreams as well as the original bit file are generated at this stage. At \circled{3}, the constraints, including the possible range for slices, regional locations, and range possible of FFs~\cite{consXilinx} in reconfiguration, are identified.
This information as well as the generated bitstreams are transferred to the online phase of the framework. Here, a lightweight Operating System (OS) (such as Ubuntu) is utilized in the SoC to generate and control the reconfiguration. A secured one-time true-random number generator (TRNG)~\cite{tsoi2003compact} is deployed in \circled{4} and the randomness is passed each time to the PR-generator unit. The PR-generator unit in \circled{5} incorporates a bitstream manipulator (Byteman\cite{manev2022byteman} in our case) with pre-defined constraints from \circled{3}. In this step, randomized LOC (the placement assignment of a logic cell in AMD FPGAs) and shuffling constraints are selected and the correspondent partial reconfiguration bitstream is generated based on the existing original bitstream from step \circled{3}. Upon generating the PR, the FPGA is programmed as the trigger signal in \circled{7} is received.

\subsection{Randomized Bitstream Manipulation}
The bitstream manipulation with the aim of one-time PR generation is the core functionality of \textit{RandOhm} online phase. The idea here is to introduce randomization in real-time rather than having the bitstreams stored in the memory as implemented in previous works \cite{khan2021moving}. This method not only increases the security level of the countermeasure but also decreases the memory utilization of the PR files to a single bitstream. As indicated in Fig.~\ref{fig:overview}, this functionality follows a simple procedure. As the TRNG unit on the processor creates a one-time randomness, the bitstream manipulator program (i.e., Byteman) collects the randomized constraint information and presents a brand new one-time bitstream. Depending on the security measurement of the target, the program signal is triggered with a specific frequency. We refer to this frequency as the \textit{PR Rate}. For the highest security level, the PR Rate is set to $1$. This means that for every single encryption process a new PR should be loaded into the target. In general, for \textit{PR Rate}$=n$, the PR regeneration is invoked after every $n$ encryption processes.


\vspace*{-5mm}
\subsection{Real-time Circuit Multiplexing}\label{sec:reg_seq}
Here, we explain how circuits can be multiplexed in real time using an example for the fine-grained reconfiguration technique discussed in Sect.~\ref{seq}. 
We incorporate DFX in \textit{RandOhm} to generate PR as a hardware-based scrambling method.
For this aim, a high-level script (e.g., python) is employed to select a random permutation of the target registers (e.g., 128 bits of AES master key). The possible search space for such permutation is super-exponential ($128!$) and, as will be shown shortly, effectively decreases the impedance leakage. This sequence is then passed to the bitstream manipulator program (i.e., Byteman) to be included as constraints in the bitstream codes. Compared to the existing solutions, this process incurs the minimum resource utilization as it only requires a single Reconfigurable Module (RM) to be implemented. This is due to the fact the only modification is the hardware referencing of the target FFs, which effectively modifies the internal local routing in the target slice, leading to randomization of the impedance profile.
 

\section{Threat Model and Attacks}

Similar to the threat models presented in~\cite{monfared2023leakyohm}, we assume both profiled and non-profiled impedance attacks under known plaintext scenarios. 
Specifically, we consider the correlation impedance attack (\textit{CIMA}) and the differential impedance attack (\textit{DIMA}) as non-profiling attacks~\cite{monfared2023leakyohm}. 
For the profiled template impedance attack (\textit{TIMA}), we assume that the random shares of a masked AES implementation, such as key shares, can be profiled.
At the execution level, the adversary measures the impedance when the target data is static between two clock cycles (by slowing or halting the clock) or when the data is at rest in certain DUT registers after the encryption is over.
This approach aligns closely with threat models of all static SCA attacks (e.g., static power analysis~\cite{moos2019static}, LLSI~\cite{krachenfels2021real}, and impedance analysis~\cite{monfared2023leakyohm}.)
From the defender prospecting, \textit{RandOhm} is deployed on the target IC, and it frequently upgrades the underlying hardware circuit using PR to prevent the aforementioned attack.
\textit{RandOhm} should be operated using an internal clock source that cannot be tampered with by the adversary.

\section{Experimental Setup}

\subsection{Measurement Equipment}\label{subsec:Measrement_setup}
\begin{figure}[t]
  \begin{subfigure}[b]{0.49\columnwidth}
    \includegraphics[width=\linewidth]{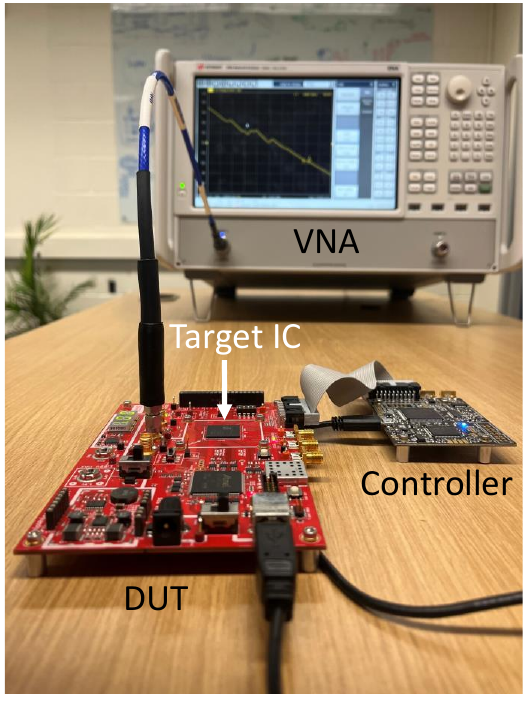}
    \caption{}
    \label{fig:experiment_set_photo}
  \end{subfigure}
  \begin{subfigure}[b]{0.4\columnwidth}
    \includegraphics[width=\linewidth]{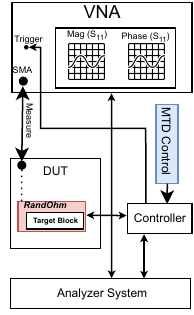}
    \caption{}
   \label{experiment_set_diagram}
  \end{subfigure}
    \caption{Measurement setup. (a) VNA capturing \textit{$S_{11}$} traces from the DUT  and (b) Experimental setup diagram.}
  \label{fig:experiment_set}
\end{figure}

In our research, we employed the Keysight ENA Network Analyzer E5080A \cite{enaKeysight}, which operates up to a 6 GHz frequency bandwidth for RF measurements. We also used Minicircuit CBL-2FT-SMNM+ shielded cables \cite{minicircuits} for scattering measurements, compatible with the same frequency bandwidth. The ports of our VNA include internal capacitors to eliminate DC voltage and, therefore, eliminate the need for a Bias Tee. Fig.~\ref{fig:experiment_set} shows the experimental setup for our evaluations.

\subsection{Device Under Test}\label{subsec:DUT}
Our experiments utilized two boards. 
For security analysis, we utilized a NewAE CW305 board~\cite{CW305}, equipped with a 28 nm AMD/Xilinx Artix-7 FPGA (XC7A100T), as it allows direct access to the FPGA's core ($V_{CCINT}$) PDN.
For overhead analysis, we used a ZedBoard AMD/Xilinx Zynq-7000 SoC Board (XC7Z020), equipped with 28 nm ARM processors and Artix-7 FPGA fabric. 



\subsection{Analyzer and Controller Configuration} 
We controlled the FPGA chip using a NewAE CW-Lite board~\cite{cwlite}, facilitating serial communication with the DUT and serving as an intermediate controller for plaintext and ciphertext transfer. The CW305 board was set up to synchronize the IC's clock with the controller's trigger signal (e.g., CW-Lite). For clock-controlled experiments (like TIMA), the target's clock signal was generated by PLLs on the CW305, with feedback sent simultaneously to the controller. Upon reaching the desired timestamp, the controller masked the target's clock signal, halting computation. Although the PLL board clock continued oscillating, the target clock on the IC was gated. This idle status triggered the VNA for measurement. We set the PLL board clock to 100 MHz. 
The \textit{Analyzer System} comprised an Intel XEON E5 2697 V3 CPU (2.6 GHz) with 128 GB RAM, running Ubuntu 20.04.6 LTS.

\subsection{Target Implementation and Configuration }
\hfill

\noindent\textbf{VNA Configurations and Frequency Bands.}
Different frequency bands were selected based on the target implementation. 
The IF Bandwidth was set to $500$ Hz to filter unwanted responses, and the \textit{Averaging factor} was $200$ in TIMA attack to minimize measurement noise. Furthermore, our analysis merely relies on the phase (quantifies by \textit{deg}) part of the impedance profile. 
\\
\textbf{Implementation of Masked AES.}
In our experiments, we focused on an AES DOM implementation~\cite{gross2016domain} with 3 shares (masking order of 2), deploying \textit{TIMA} attacks. We utilized AES DOM \textit{VHDL} reference code with a wrapper. The measurements targets the first round's key-share byte registers before the S-box operation. For \textit{DIMA} and \textit{CIMA}, we consider the first byte of key from the first byte S-box output in the state register of an unprotected AES implementation.

\noindent\textbf{MTD Implementation.}
For security analysis, the PRs are generated on a \textit{Analyzer System} as it is directly connected to the NewAE CW305 (Artix-7 FPGA) board with a serial connection. 
These experiments were conducted with a PR generation rate of 16 using \textit{RandOhm}.
For overhead analysis, we deploy \textit{RandOhm} on the ZedBoard (Zynq-7000 ARM/FPGA) and utilize the ARM cores to implement the real-time PR generation, as described in Sect.~\ref{modular}. 
In the latter experiment, \textit{RandOhm} is operated on a Petalinux 2019.2 kernel loaded onto an external 8GB SD Card. Moreover, for the reconfiguration process, we employed the Internal Configuration Access Port (ICAP) interface~\cite{artix7}. 


\begin{figure}[t]
  \begin{subfigure}[b]{0.49\columnwidth}
    \includegraphics[width=\linewidth]{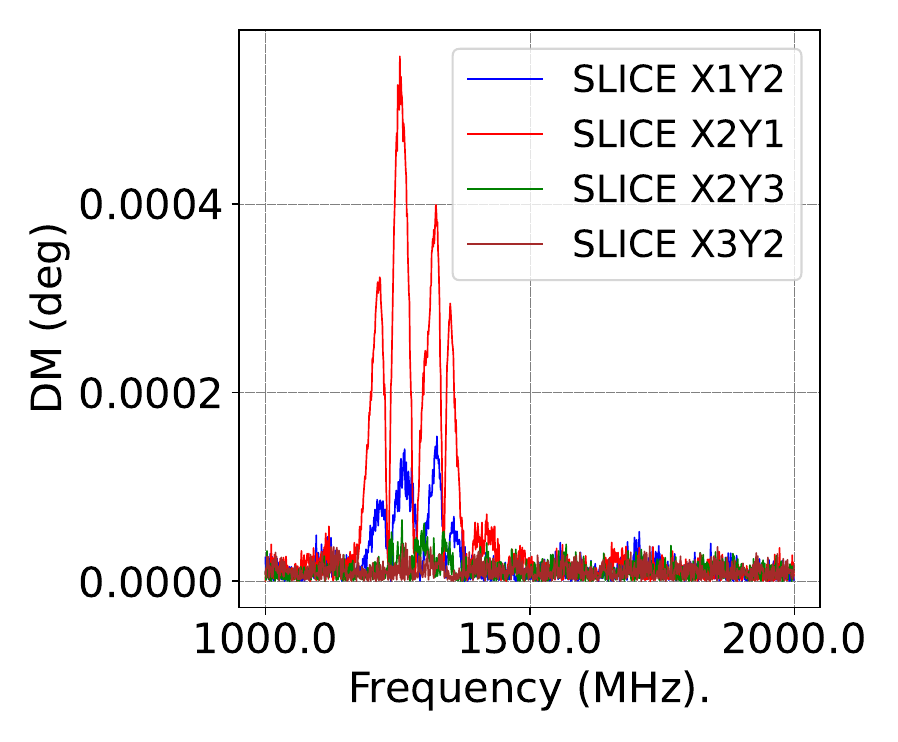}
    \caption{}
    \label{fig:imp_init_a}
  \end{subfigure}
  \hfill 
  \begin{subfigure}[b]{0.49\columnwidth}
    \includegraphics[width=\linewidth]{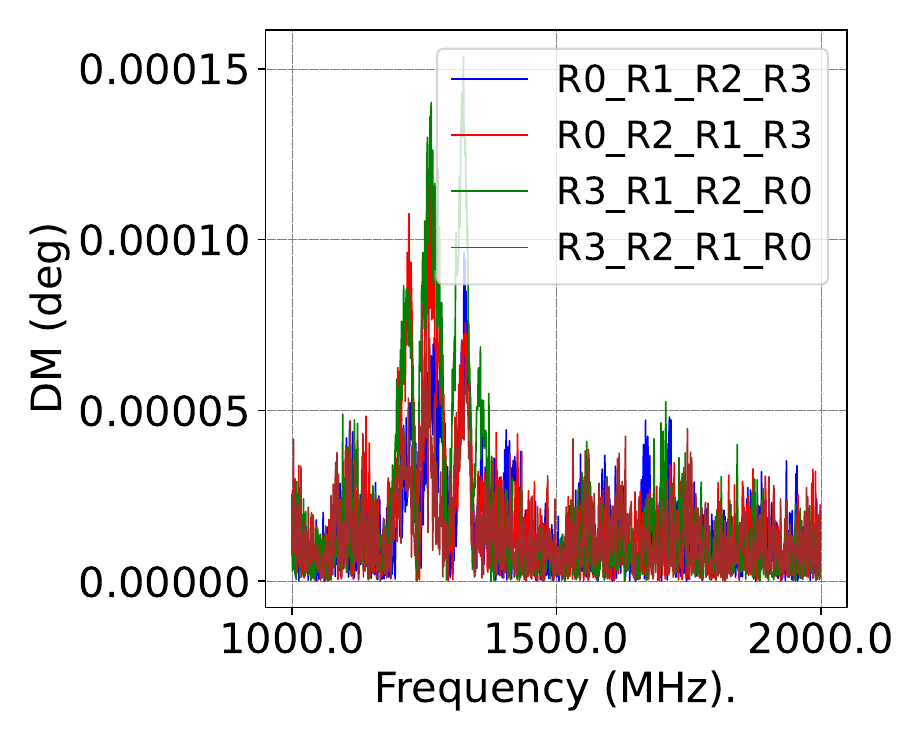}
    \caption{}
   \label{fig:imp_init_b}
  \end{subfigure}
    \caption{Impedance template leakage on proposed hardware multiplexing (Difference of Means for each target bit). (a) Target Register Location Multiplexer (b) Target Register Sequence Multiplexer.}
  \label{fig:imp_init}
\end{figure}

\begin{figure*}[!t] 
\centering
\includegraphics[width=0.9\linewidth]{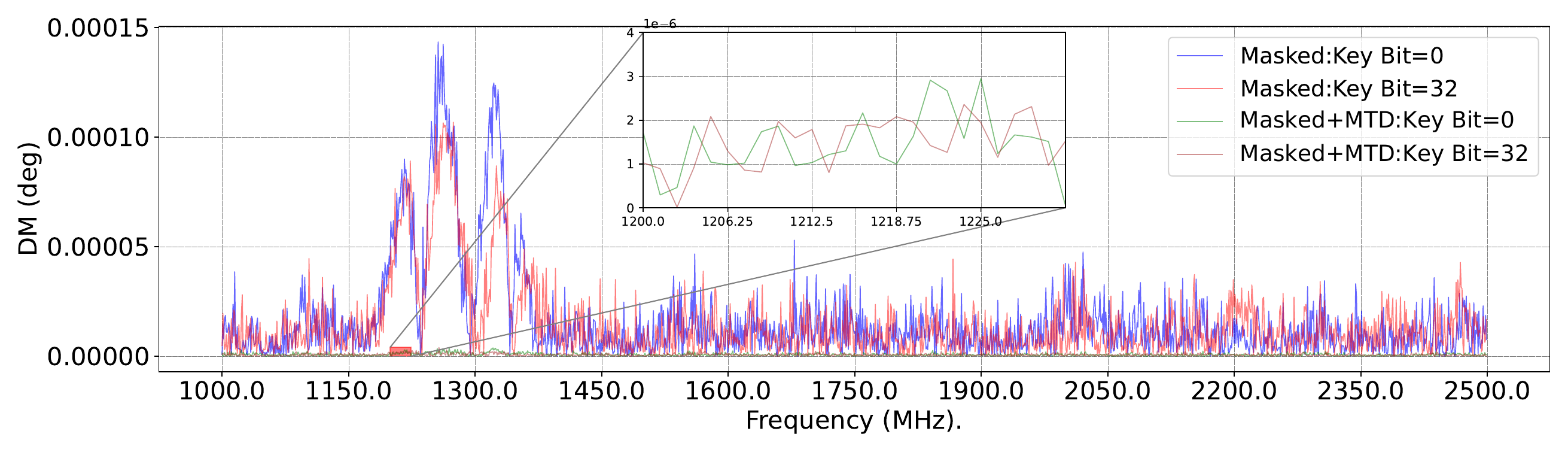}
\caption{Illustration of TIMA leakage of AES DOM for two target masked bits in presence of MTD }
\label{fullkey}
\end{figure*}

\section{Results}

\subsection{Initial Observations}
To illustrate the effectiveness of the described methods, the results of a template attack via impedance analysis~\cite{monfared2023leakyohm} are provided in this section. Although we will perform a detailed  analysis with regards to MTD against Template Impedance Analysis (TIMA), here we only showcase initial observations with regards to measured impedance leakage when employing the proposed MTD.

Fig.~\ref{fig:imp_init} demonstrates the impedance template leakage on both described hardware multiplexing in Sect.~\ref{sec:multiplex}. Fig.~\ref{fig:imp_init_a} shows the template impedance leakage on a single target bit where slice multiplexing is activated and the target bit is loaded into four different slices. As shown, a considerable amount of frequency shift as well as leakage change is measured for this scenario. On the other hand, Fig.~\ref{fig:imp_init_b} shows the impedance leakage of the same bit where it is implemented in the same slice but scrambled with adjacent registers. As depicted, the frequency shift, as well as leakage variations in this case is much less compared to the slice-swapping technique. However, since TIMA is a bit-wise attack, small single-bit alterations in leakage, significantly reduce the success rate during the attack phase~\cite{monfared2023leakyohm}. 

For the rest of the paper, we deploy target register multiplexing and show that it resists impedance attacks.

\subsection{Profiled Attack}
Here, we perform TIMA~\cite{monfared2023leakyohm} against the masked AES implementation. 
We initiate the attack at the first clock cycle when the shares of the first key byte and the first input byte shares are loaded into the target. This approach allows TIMA to directly attack the key (share) registers, bypassing any masked operations in subsequent clock cycles.
During the profiling phase, we conduct two sets of experiments in \textit{1GHz-3GHz} frequency range with 5000 frequency points: 1) we collect 10,000 traces to create templates for the masked key registers without utilizing \textit{RandOhm}. 2) Then we activate \textit{RandOhm} and perform the profiling stage for 100,000 traces.
TIMA profiling is conducted independently for each bit of all key shares. 
Note that the shares are generated in a uniformly random manner, and each trace contributes to template of all target bits. 
Specifically, for templating each target bit, we approximately have 10,000 traces where a target bit of the concerned share is 0b0 in the first scenario. After the profiling stage, the adversary attempts to guess the key based on a limited number of attack traces. Here, to analyze the information leakage, we use a simple Difference of Mean (DM) metric. This means that for a specific profiled target key bit we have: ($DM_{b_t} = Abs(Mean(Tr|b_t=0b0)-Mean(Tr|b_t=0b1))$). If $DM$ is large enough to be distinguished (in terms of relative SNR), the adversary can effectively perform the template attack.

Fig.~\ref{fullkey} shows the the impedance DM leakage for two target bits of masked keys in frequency domain. It is clearly observed that \textit{RandOhm} contributes to impedance profile randomization. Note that DM leakage is averaged over 100,000 traces for MTD-enabled implementation compared to 10,000 traces in the regular AES-DOM implementation.

\subsection{Non-Profiled Attacks}
\begin{figure}[t]
  \begin{subfigure}[b]{0.49\columnwidth}
    \includegraphics[width=\linewidth]{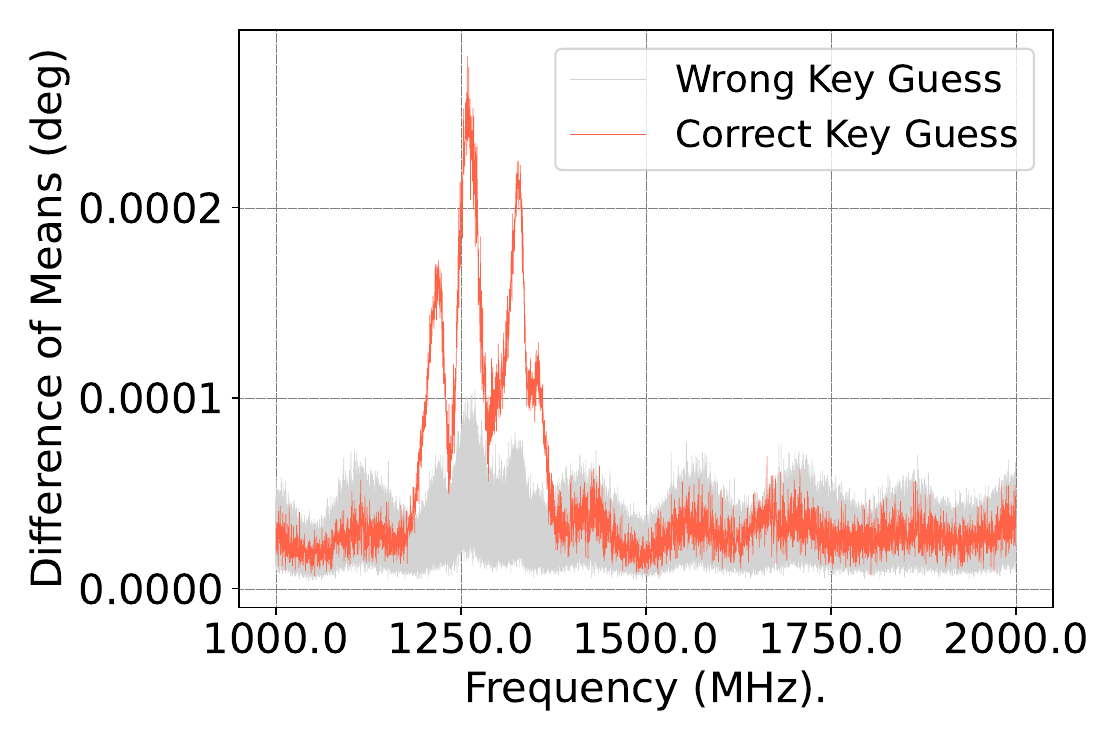}
    \caption{}
    \label{fig:dima_leak_a}
  \end{subfigure}
  \hfill 
  \begin{subfigure}[b]{0.44\columnwidth}
    \includegraphics[width=\linewidth]{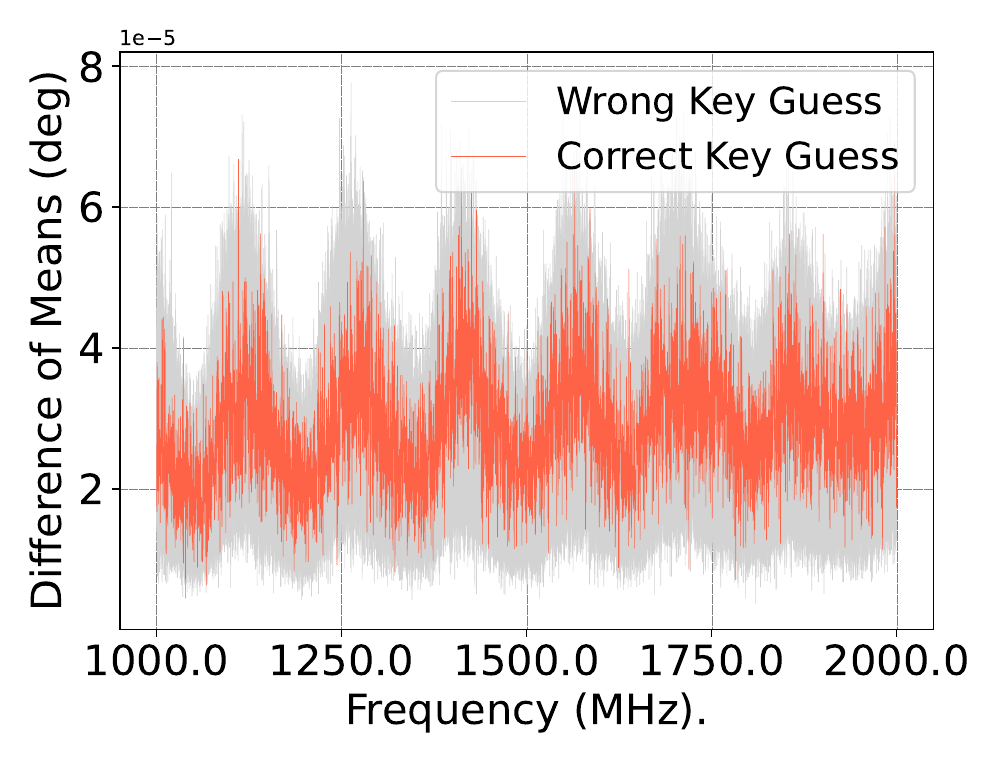}
    \caption{}
   \label{fig:dima_leak_b}
  \end{subfigure}
    \caption{DIMA Attack for $N=10,000$ traces on first byte key. (a) Without using \textit{RandOhm} (b) With \textit{RandOhm}.}
  \label{fig:dima_leak}
\end{figure}

\noindent\textbf{Differential Impedance Analysis.} In our subsequent attack scenario, we conducted two DIMA attacks on an AES S-Box with and without the presence of \textit{RandOhm}. These attacks involved analyzing 10,000 traces within the frequency band of 1 GHz to 2 GHz, with each measurement comprising 3000 frequency samples. The comparison differential results for the potential key space, derived from a multi-bit DIMA analysis, are illustrated in Fig.~\ref{fig:dima_leak}.
Additionally, we explored and measured the leakage with respect to the number of traces used for both scenarios. While some fluctuations appear in the presence of \textit{RandOhm}, the DM leakage tends to stay indistinguishable as the number of attack traces increases.

\begin{figure}[t]
  \begin{subfigure}[b]{0.49\columnwidth}
    \includegraphics[width=\linewidth]{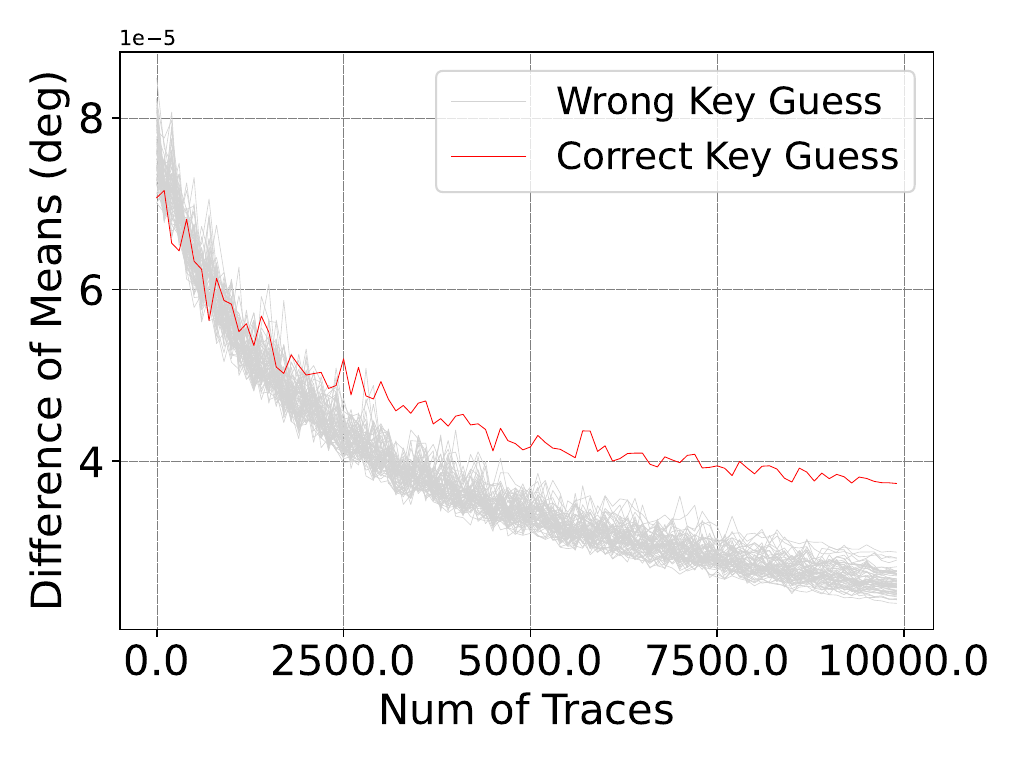}
    \caption{}
    \label{fig:dima_leak_trace_a}
  \end{subfigure}
  \hfill 
  \begin{subfigure}[b]{0.49\columnwidth}
    \includegraphics[width=\linewidth]{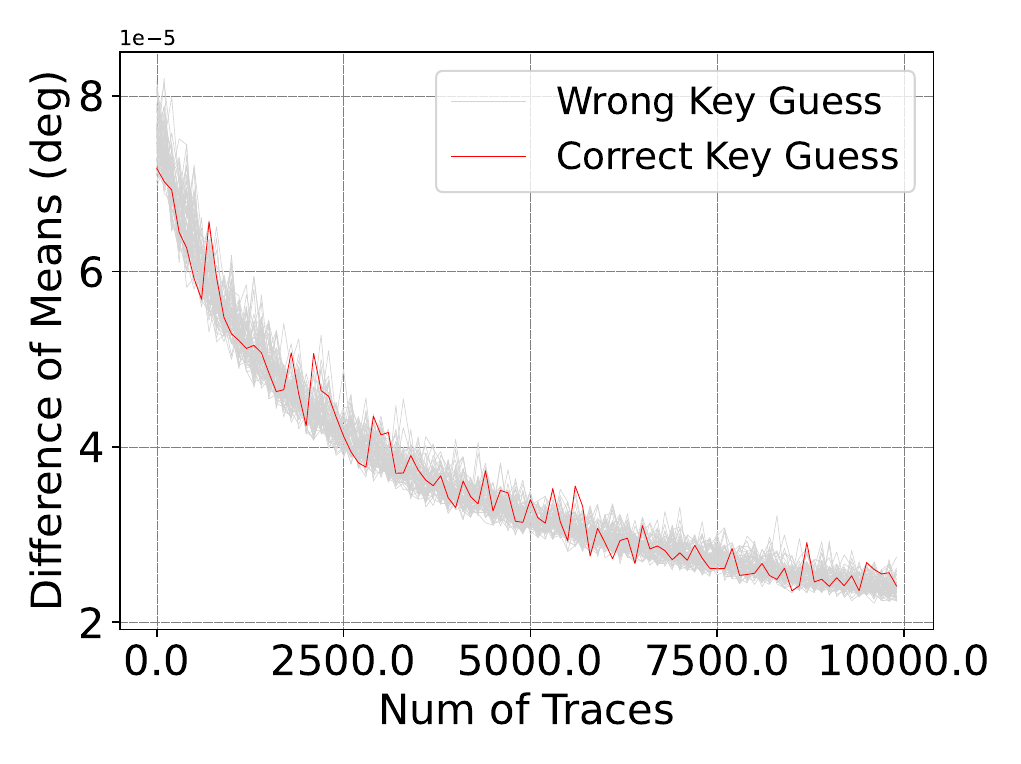}
    \caption{}
   \label{fig:dima_leak_trace_b}
  \end{subfigure}
    \caption{DIMA Attack for different traces on first byte key. (a) Without using \textit{RandOhm} (b) With \textit{RandOhm}.}
  \label{fig:dima_leak_trace}
\end{figure}

\begin{figure*}[t]
  \begin{subfigure}[b]{0.47\textwidth}
    \includegraphics[width=\linewidth]{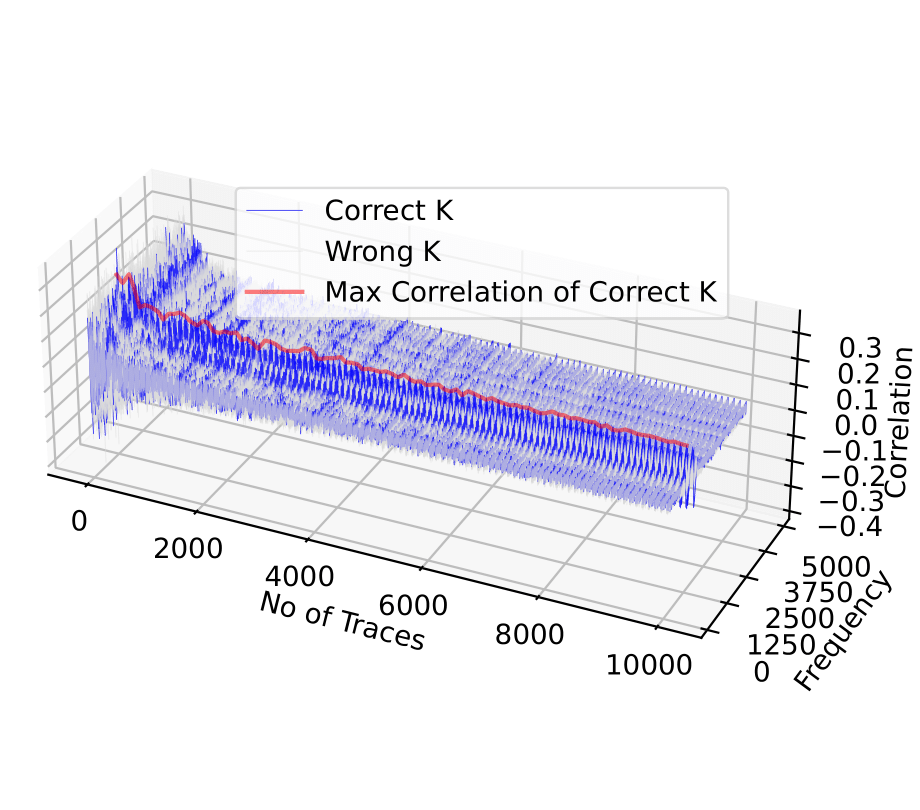}
      \vspace{-15mm}

    \caption{}
    \label{fig:cima_trace_a}
  \end{subfigure}
  \begin{subfigure}[b]{0.47\textwidth}
    \includegraphics[width=\linewidth]{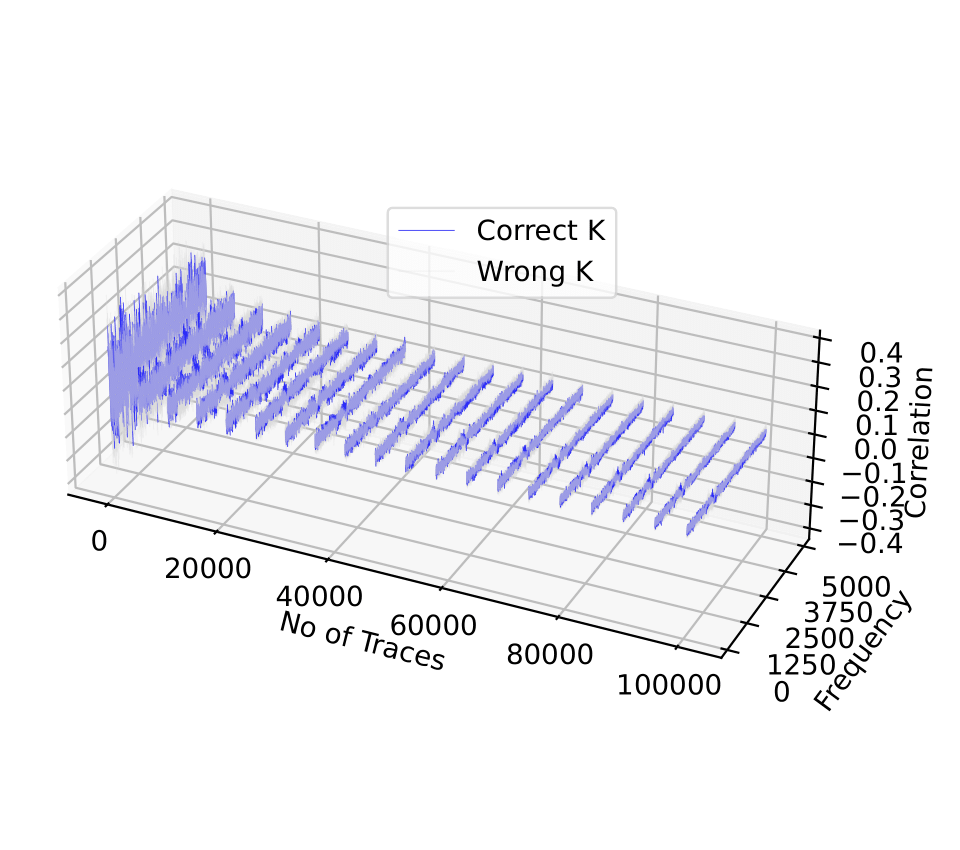}
      \vspace{-15mm}

    \caption{}
    \label{fig:cima_trace_b}
  \end{subfigure}
    \caption{Leakage measurement of CIMA with respect to number of traces 
 in frequency domain. (a) Without using \textit{RandOhm} (b) With \textit{RandOhm}.}
    \label{fig:cima_trace}
\end{figure*}

\begin{figure}[!h]
  \begin{subfigure}[b]{0.49\columnwidth}
    \includegraphics[width=\linewidth]{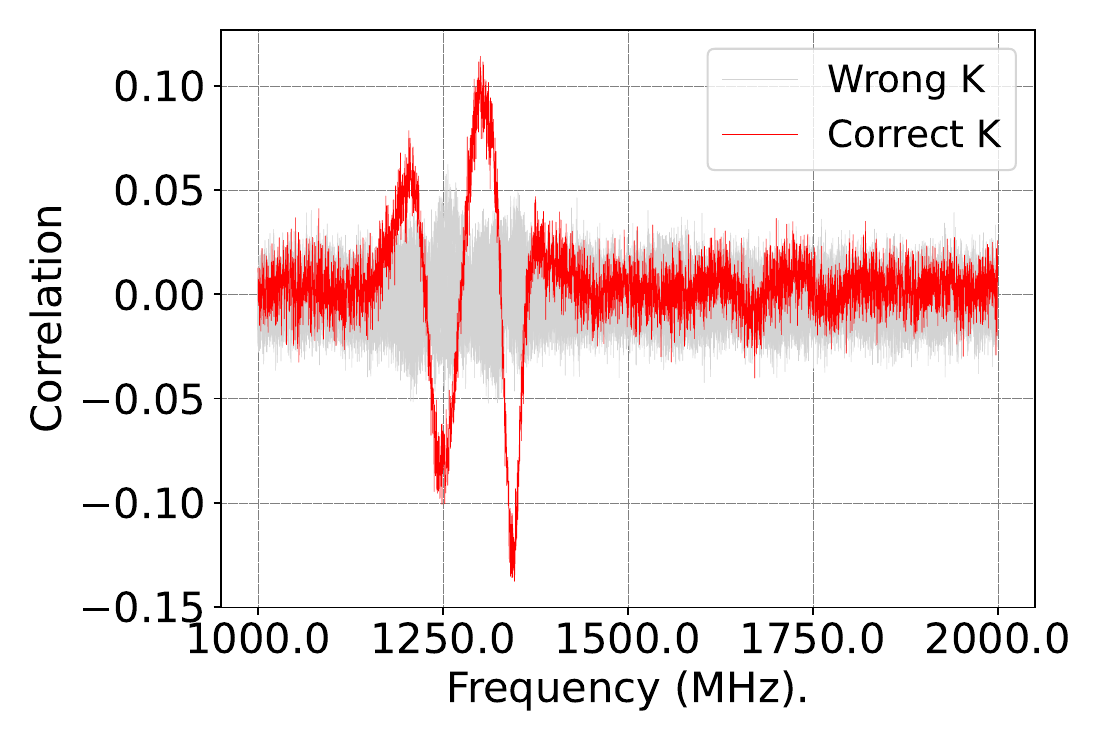}
    \caption{}
    \label{fig:cima_leak_a}
  \end{subfigure}
  \hfill 
  \begin{subfigure}[b]{0.49\columnwidth}
    \includegraphics[width=\linewidth]{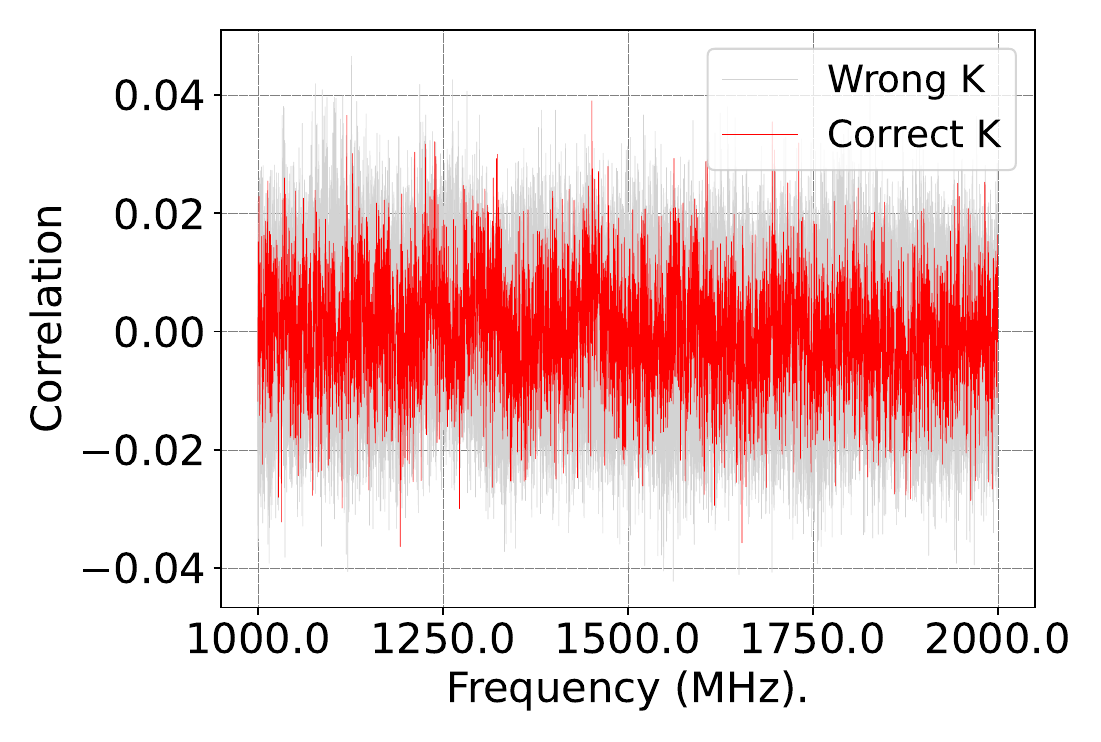}
    \caption{}
   \label{fig:cima_leak_a}
  \end{subfigure}
    \caption{CIMA Attack for $N=10,000$ traces on first byte key.(a) Without using \textit{RandOhm} (b) With \textit{RandOhm}.}
  \label{fig:cima_leak}
\end{figure}

\begin{figure}[!h]
  \begin{subfigure}[b]{0.49\columnwidth}
    \includegraphics[width=\linewidth]{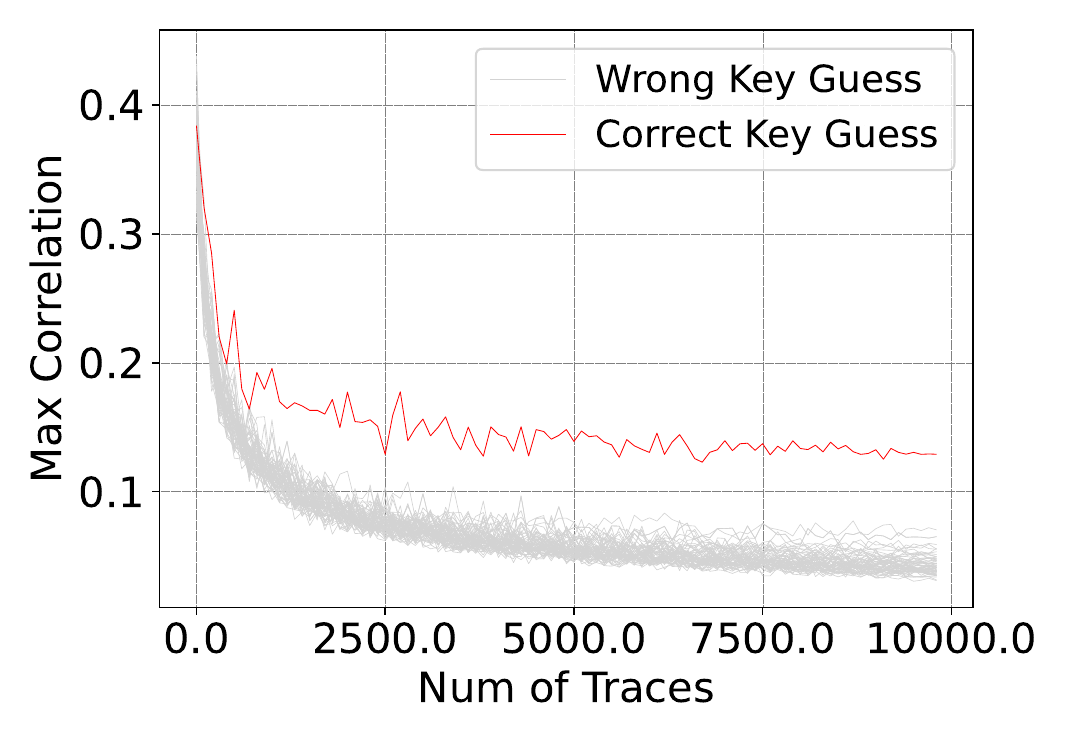}
    \caption{}
    \label{fig:cima_leak_trace_a}
  \end{subfigure}
  \hfill 
  \begin{subfigure}[b]{0.49\columnwidth}
    \includegraphics[width=\linewidth]{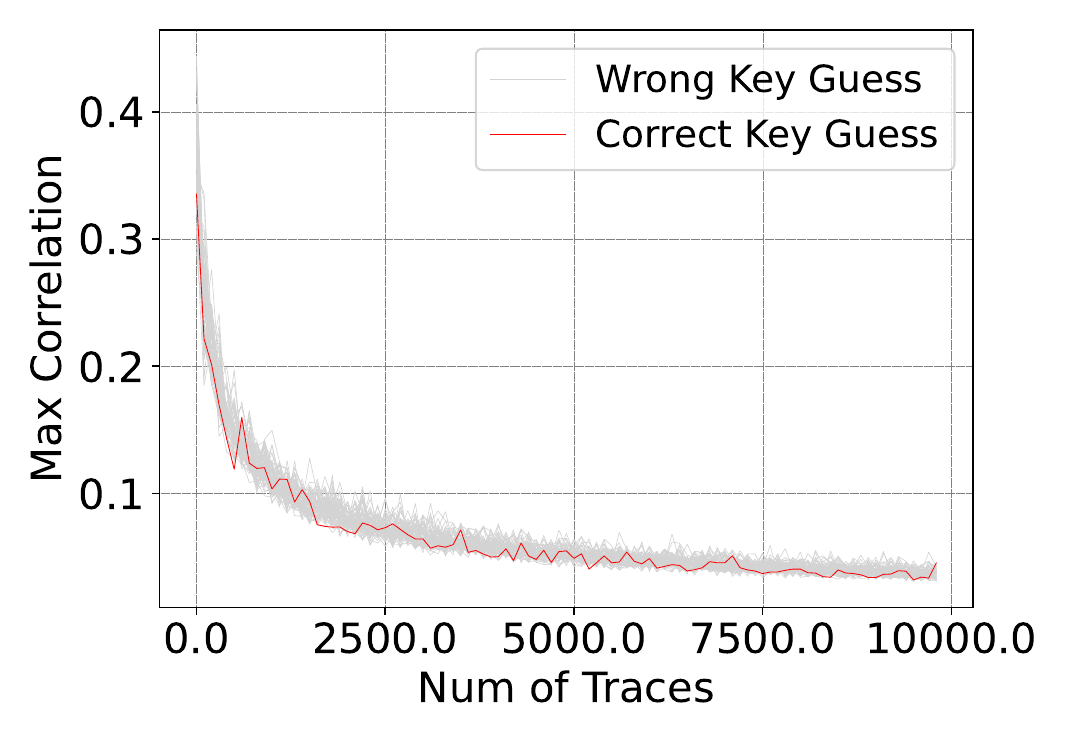}
    \caption{}
   \label{fig:cima_leak_trace_b}
  \end{subfigure}
    \caption{CIMA Attack for different traces on first byte key. (a) Without using \textit{RandOhm} (b) With \textit{RandOhm}.}
  \label{fig:cima_leak_trace}
\end{figure}

\noindent\textbf{Correlation Impedance Analysis.} In a further attempt to break AES using impedance data and evaluate \textit{RandOhm}, we conducted CIMA attacks using a standard Hamming Weight (HW) model. This experiment involved analyzing 10,000 measurement traces, focusing on a frequency range of 2GHz to 3GHz, with 3000 linearly distributed frequency points. The correlation index results from CIMA attack are presented in Fig.~\ref{fig:cima_leak}.
Moreover, to confirm the effectiveness of the proposed mitigation we also perform correlation analysis based on the number of traces. Fig.~\ref{fig:cima_leak_trace} shows the progressive maximum correlation of the correct key up to 10,000 traces for each of the experiments. Similarly, it is observed that correlation leakage does not increase to the limit of 10,000 traces.

To evaluate \textit{RandOhm} against CIMA, we increase the number of traces to 100,000 traces and perform a similar analysis when \textit{RandOhm} is activated. Fig.~\ref{fig:cima_trace} depicts the progressive maximum correlation in the frequency domain as the number of used traces increases. As highlighted in Fig.~\ref{fig:cima_trace_a}, the correct key shows a maximum correlation in multiple frequency points (highlighted in dark blue), where the proposed MTD strategy mitigates the information leakage exploited by correlation in the frequency domain.

\vspace*{-4mm}
\subsection{Overhead Analysis}
Here, we consider the implementation of the \textit{RandOhm} on the Zed-board ARM/FPGA Zynq-7000 SoC.
For delay overhead, the offline part of the \textit{RandOhm} process is not considered, as it is executed only once during the design.
The online part comprises a real-time randomized PR generation on the ARM cores in the SoC, which incurs a constant delay for each bitstream generation. As indicated, in this part we use Byteman as a fast PR bitstream generator instead of regular AMD/Xilinx pipeline PR generator. The final part is the bitstream loading which is carried out by ICAP interface. For non-profiled and profiled scenarios, we consider state registers and key register protection, respectively. Using the ICAP at 100MHz with a 400MB/s data transfer rate, the PR loading procedure takes approximately $74$ clock cycles for each CLB~\cite{tang2020partitioning} for inner-CLB register swapping reconfiguration if the registers (up to 16 FFs) in a single CLB are protected.


The masked AES-DOM implementation in our evaluations, without using \textit{RandOhm}, utilizes 7426 LUTs and 3581 FFs.
The overhead for the register shuffling method described in Sect~\ref{sec:reg_seq} is negligible ($<0.1\%$) since it is executed within the same RM.
In the case of coarse grain randomization (which was not used in this paper to protect the AES), the overhead~\cite{ahmadi2023fpga,khan2021moving} in AES circuits will be up to 14\% in terms of number of LUTs and FFs.

  \section{Discussions}\label{sec:dis}


\noindent\textbf{Speeding up \textit{RandOhm}}. It is possible to parallelize the reconfiguration process alongside with the target encryption to further improve the throughput of the \textit{RandOhm}. Authors in \cite{khan2021moving} argue that if the generated partial bitstreams are small in size, the ICAP interface is capable of programming the device trivially in a way that does not incur any bottleneck for the encryption. However, the triggering mechanism should be carefully managed to prevent any malfunctioning of the target algorithm. In the case of \textit{RandOhm}, if key registers are targeted to mitigate impedance attacks, this could effectively be done after the key-scheduling process. We leave this implementation and related considerations for future work.

\noindent\textbf{Integration with Side-Channel Sensors.} Previous researchers have shown that powerful side-channel threat models could potentially damage or disable the countermeasure mechanism. These side-channels are required to be detected before being mitigated \cite{cannon2023protection}. \textit{RandOhm} only performs as an active countermeasure in the system. Although implemented efficiently, similar to other hardware countermeasures \textit{RandOhm} also incurs resource utilization and delay overhead. One can argue that a system-level detection sensor (e.g.,~\cite{mosavirik2022impedanceverif}) for physical attacks, including impedance analysis, could be implemented which can serve as a trigger to activate \textit{RandOhm}. This will increase the efficiency of the mitigation by decreasing the overhead and also could be considered as a hidden mitigation mechanism for some applications.



 \vspace{-4mm}
\section{Conclusion}\label{sec:Conclusion}
In this paper, we put forward a new technique called \textit{RandOhm} that leverages MTD principles through the PR feature of conventional FPGAs. 
We conducted a comprehensive study on the sources of impedance leakage inside the FPGA fabric and showed that such reconfigurations can thwart impedance side-channel attacks by regularly randomizing the placement and routing of sensitive circuits.
By deploying open-source bitstream manipulators, we built a real-time PR-based countermeasure for programmable SoCs/FPGAs.
We demonstrated the resiliency of our approach by mounting impedance attacks against the implementation of AES ciphers using \textit{RandOhm} on 28 nm FPGAs.
Based on the results of our AES-DOM implementation, we showed that \textit{RandOhm} is transparent to other algorithmic countermeasures, such as masking, and can be combined with them to resist both passive and backscattered SCA attacks.
Finally, we examined the overhead of our proposed scheme in terms of delay and resource utilization.

\section*{Acknowledgment}
This effort was sponsored by NSF Grant CNS-2338069.

\bibliographystyle{ACM-Reference-Format}
\bibliography{ref}

\end{document}